# Lung image segmentation by generative adversarial networks

Jiaxin Cai*, Hongfeng Zhu

School of Applied Mathematics, Xiamen University of Technology, Xiamen, P.R. China

## ABSTRACT

Lung image segmentation plays an important role in computer-aid pulmonary diseases diagnosis and treatment. This paper proposed a lung image segmentation method by generative adversarial networks. We employed a variety of generative adversarial networks and use its capability of image translation to perform image segmentation. The generative adversarial networks was employed to translate the original lung image to the segmented image. The generative adversarial networks based segmentation method was test on real lung image data set. Experimental results shows that the proposed method is effective and outperform state-of-the art method.

## 1. INTRODUCTION

Machine learning[1, 2] has received lots of research interest due to its wide application fields such as computer vision, speech recognition[3], and natural language processing[4]. Recently, deep learning has become the most popular research topic in the field of machine learning. More and more people have focused on deep learning based image analysis, including image processing [5], image generation[6] and video retrieval [7]. Generative Adversarial Networks (GAN) [8] is a popular deep learning model in the research field of computer vision. It has become one of the most valuable technologies for image generation.

Medical image segmentation plays an important role in computer-aid clinical diagnosis and treatment. Traditional medical image segmentation algorithms mainly include banalization methods[9], watershed algorithms[10], level set based methods[11], and stochastic optimization models[12]. However, these algorithms achieve good results only when there is a large difference between the background area and the object area. In order to improve the accuracy of segmentation, curvature constraints and local feature constraints are often added into the models. However, these methods are still not effective when the object area is similar to background. Therefore, using deep learning algorithm to solve medical image segmentation has always been a research direction with theoretical significance and practical application value. Deep learning uses multi-layer networks and learns rich local features. It has strong fitting ability, but also has some shortcomings such as time-consuming training process and difficulty of explaining concretely. Deep learning based medical image segmentation usually adopts full convolution neural network (FCN) [13], Mask R-CNN[14], U-Net[15], and other deep networks. The advantages of these algorithms are that they have multi-layer structures to generate rich features which are helpful to improve the recognition performance. The disadvantages of these algorithms are that the training time is often very long, and the segmentation result is still not fine enough[16]. Sometimes it cannot be guaranteed that the segmentation effects of them have a great improvement than traditional algorithms.

Lung images have the characteristics of multimodality, gray level fuzziness and uncertainty, which lead to unsatisfactory segmentation effect for many algorithms. Specially, pulmonary nodules are very small in size. Some nodules are even hidden. Once the segmentation details are not well processed, these nodules are hard to detect by tradition deep learning models. To solve this problem, this paper proposed a segmentation method based on GAN. We employed the Pix2Pix[17] networks, a variety of GAN, to adopt lung images. Pix2Pix is an image translation algorithm that can transform blurred image to exact image. We employed Pix2Pix to translate the lung image to the segmented image. We took the original lung image as blurred image, and took the segmented image as exact image. Then we translated the blurred image into exact image by Pix2Pix. The result of translation was taken as the segmentation result. The Pix2Pix segmentation method was test on lung image data set. Experimental results demonstrate that this method is effective and has better performance than state-of-the art method.

*caijiaxin@xmut.edu.com

## 2. METHOD

### 2.1 GAN

GAN[8] is a network developed by Goodfellow et al. in 2014. GAN employs an algorithm that achieves the optimization goal through repeated antagonistic games[16]. It consists of a generator and a discriminator. The main purpose of the generator is to generate enough false images so that the discriminator cannot determine whether it is true or not. The purpose of the discriminator is to ensure that they are not deceived by the generator. The classical generative models usually need to give a specific framework and require parameter estimation which needs complex calculations such as Monte Carlo sampling or other approximate estimation procedures. Different to classical generative models, GAN does not need complex calculation about probability. GAN does not need to specify the distribution type. It directly simulates the distribution using real data by deep neural networks. The gradient descent algorithm based on BP algorithm is usually used in the training process. The optimization function of GAN can be written as follows:

$$\min_G \max_D V(D,G) = E_{x \sim p_{data}(x)} \log D(x) + E_{z \sim p_z(z)} \log(1 - D(G(z))) \tag{1}$$

where $G$ is the generator, and $D$ is the discriminator. $x$ is the real picture. $z$ is the noise input of the generator $G$. $G(z)$ is the picture generated by generator $G$. $D(x)$ is the probability that the discriminator D judges whether the real picture is true. $D(G(z))$ is the probability that the discriminator judges whether the picture generated by the generator G is true or not.

The purpose of GAN is to learn the distribution of training data[16]. In order to accomplish this goal, first of all, a noise is input into the generator. The generator transforms this noise into a picture. The discriminator identifies these simulated pictures with the real picture, and gives the true and false coefficients of the image. Through cyclic alternate training, the generation and the discriminator are both improved. The generator can generate synthetic images which are very similar to the original images[16].

### 2.2 Pix2Pix

Pix2Pix[17] is a framework developed based on conditional GAN (cGAN) [18]. Similarly, Pix2Pix has a generator G and a discriminator $D$. The input and output of $G$ are both a single image. In order to ensure that the generated image matches the input image, the loss function of conditional GAN takes the form as follows.

$$\Gamma_{cGAN}(G,D) = E_{x,y}\left[\log D(x,y)\right] + E_{x,z}\left[\log(1 - D(x, G(x,z)))\right] \tag{2}$$

where $G$ is the generator and $D$ is the discriminator. $z$ is the input random vector. $x$ is the image to be converted. $y$ is the target image. $\Gamma_{cGAN}$ is the conditional GAN loss function.

In the process of image translation, a lot of information is shared between the input and the output of the generator $G$. So an *L1* loss is added to ensure the similarity between the input image and the output image is large.

$$\Gamma_{L1}(G) = E_{x,y,z}\left[\|y - G(x,z)\|_1\right] \tag{3}$$

Finally, the loss function of Pix2Pix is constructed based on merging the conditional GAN loss function and the L1 loss function. The solution of Pix2Pix is written as follows.

$$G^* = \arg \min_G \max_D \Gamma_{cGAN}(G,D) + \lambda \Gamma_{L1}(G) \tag{4}$$

where $\Gamma_{cGAN}(G,D)$ denotes the conditional GAN loss function, and $\Gamma_{L1}(G)$ denotes the added *L1* loss function. G* denotes the final solution.

### 2.3 Lung image segmentation by Pix2Pix

Pix2Pix is a model that can transform blurred image to exact image. We employed Pix2Pix to translate the original image to the segmented image. We took the original gray image of lung as blurred image, and made the segmented image as exact image. Then we translated the blurred image into exact image by Pix2Pix. The result of translation was taken as the segmentation result.

# 3. EXPERIMENTAL RESULTS AND ANALYSIS

## 3.1 Experimental environments

The experimental environments were Windows 10 OS, CPU i5-4210U @ 1.70GHz, 8GB memory, Anaconda 3 (64-bit), Spyder 3.3.3, Numpy package, Pillow package, Pytorch 0.4.0 package, and Torchvision 0.2.1 package. No GPUs were used for experiments.

## 3.2 Data set

We chose the data set "Finding_lungs_in_CT_data" [19] for testing the methods. The data set was published by Kevin Mader. The last update time was 2017, and the download time was March 2019. It had a set of hand-segmented lungs images. There are 2D and 3D images in the data set. In this study, we did not use the 3D images, and only tested the 2D images. There are 267 2D original gray-scale images, and 267 lung segmentation images corresponding to them one by one. 237 images of the lung images were used as the training set. The examples of training images and there corresponding ground true images are shown in Figure 1. The other 30 lung images were used as the test set. The examples of test images and there corresponding ground true images are shown in Figure 2.

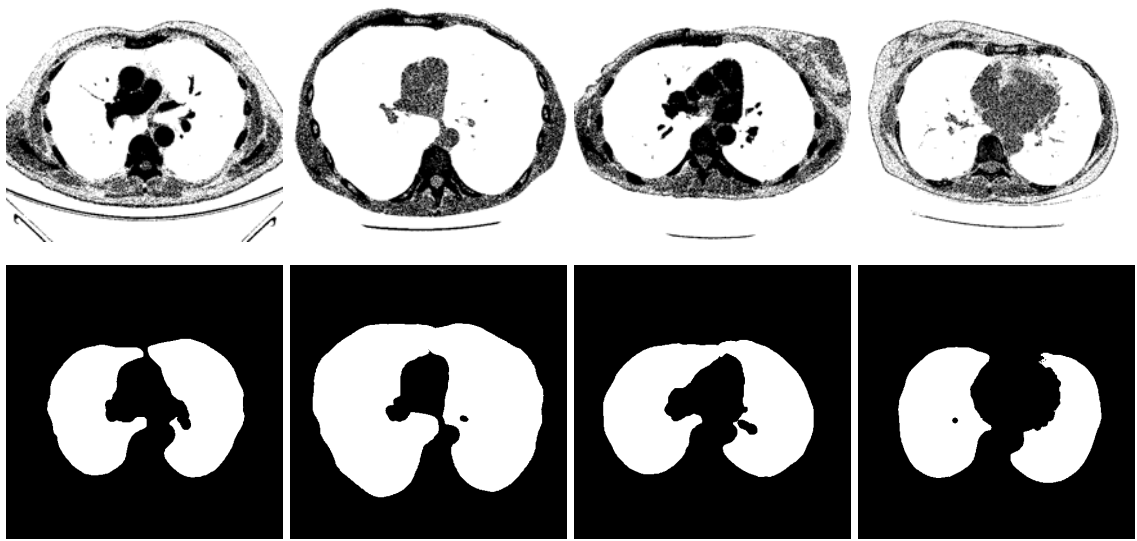

Figure 1. The examples of training images and there corresponding ground true images

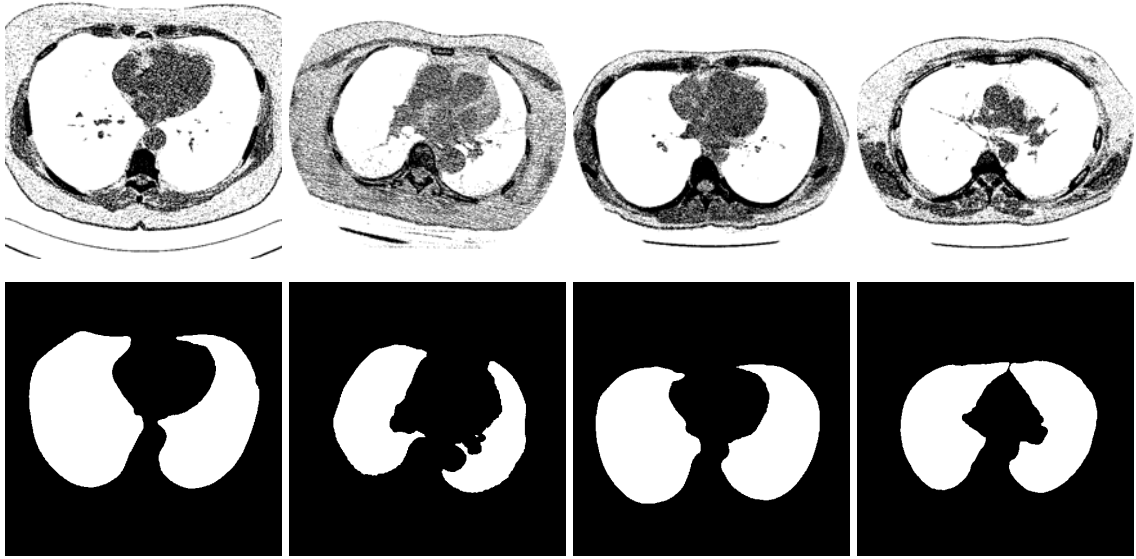

Figure 2. The examples of test images and there corresponding ground true images

### 3.3 Performance evaluation

The segmentation result of test images were compared with the ground true image which are manually segmented by experts. The accuracy, overlap rate and F measure were calculated. The computation formulas are as follows:

$$Accuracy = (TP + FN) / (TP + TN + FP + FN) \tag{5}$$

$$Overlap\ rate = |A \wedge B| / |A \vee B| \tag{6}$$

$$F = 2P*R / (P + R) \tag{7}$$

where *TP* is the true positive rate, and *FP* is the false positive rate. *TN* is the true negative rate, and *FN* is the false-negative rate. *A* denotes the segmentation result of Pix2Pix, and *B* denotes the image segmented manually by experts. *P* denotes the precision, and *R* denotes the recall. *F* denotes the F measure.

### 3.4 Experimental results

Table1 shows the experimental results of Pix2Pix based lung segmentation. The training loop was 100. As can be seen from Table1, the average accuracy of Pix2Pix based lung segmentation is 93.40%. The range of accuracy is [56.13%, 96.00%], and the standard deviation of accuracy is 7.08%. The average overlap of Pix2Pix based lung segmentation is 91.69%. The range of overlap is [32.67%, 97.54%], and the standard deviation of overlap is 13.04%. The average F measure of Pix2Pix based lung segmentation is 95.03%. The range of F measure is [49.26%, 98.75%], and the standard deviation of F measure is 9.67%. Figure 3 shows some examples of test images and there experimental results. From Figure 3, we can see that the segmentation results of Pix2Pix approximate the ground true. Experimental results demonstrate that our proposed method is effective and achieved considerable performance.

Table 1. The experimental results of Pix2Pix based lung segmentation

|  | minimum | maximum | mean | standard deviation |
|---|---|---|---|---|
| accuracy | 0.5613 | 0.9600 | 0.9341 | 0.0708 |
| overlap rate | 0.3267 | 0.9754 | 0.9169 | 0.1304 |
| F measure | 0.4926 | 0.9875 | 0.9503 | 0.0967 |

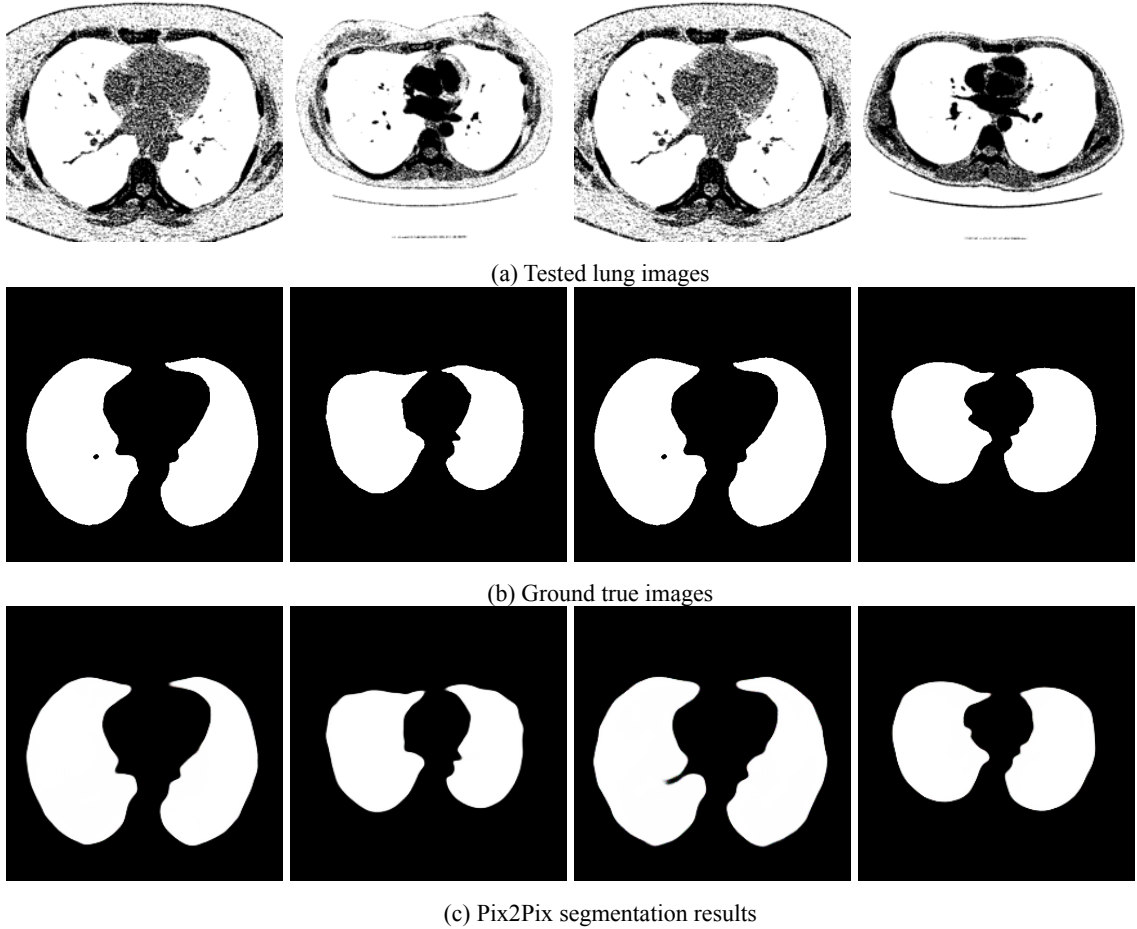

(a) Tested lung images

(b) Ground true images

(c) Pix2Pix segmentation results

Figure 3. The examples of test images and there experimental results

### 3.5 Comparison to the state-of-the-art method

U-Net was chosen as the comparison algorithm. We compared the Pix2Pix based segmentation to U-Net based segmentation on the lung data set. 67 samples in the training set were used for training, and 30 samples in the test set were tested. The training loop was 20. The accuracy with standard deviation, the overlap rate with standard deviation, the F measure with standard deviation of the two algorithms were compared. We have also compared the training time and the test time of the two methods. The comparison results are shown in Table 4. Compared to U-Net, Pix2Pix has better segmentation accuracy, overlap rate and F measure. The comparison results shows Pix2Pix has better segmentation effect. Note that using 67 samples for training was far from achieving the best performance for Pix2Pix. The Pix2Pix took half of the training and test time of U-Net, which indicates that Pix2Pix algorithm is more time-efficient than U-Net.

Table 2. The comparison results between Pix2Pix and U-Net

|  | Accuracy | Overlap rate | F measure | Training time | Test time |
|---|---|---|---|---|---|
| Pix2Pix | 0.835 ±0.081 | 0.786 ±0.157 | 0.871 ±0.115 | 6 hours | 2min25s |
| U-Net | 0.820 ±0.070 | 0.690 ±0.109 | 0.811 ±0.085 | 12 hours | 6min12s |

## 4. CONCLUSION

This paper proposed a lung segmentation method using Pix2Pix. The Pix2Pix was employed to translate the original lung image to the segmented image. The Pix2Pix segmentation method was test on real lung image data set. Experimental results shows that the proposed method is effective and outperform state-of-the art method.

## ACKNOWLEDGMENT

The authors would like to thank the Natural Science Foundation of Fujian Province (No.2016J01040) and the Xiamen University of Technology High Level Talents Project (No.YKJ15018R).